
\magnification 1200
\overfullrule=0pt
\def\gta{\mathrel{\vcenter{\offinterlineskip\halign{\hfil
$\textstyle##$ \hfil\cr>\cr\sim\cr}}}}

\def\eV{\hbox{\rm eV} }
\def\keV{\hbox{\rm keV} }

\def\nuc.#1.#2.{\ifmmode \sp{#2}{\rm#1} \else $\sp{#2}{\rm#1}$ \fi}

\baselineskip=14pt
\null\vskip 2.0cm
\centerline{\bf Experimental nuclear masses and the ground state}
\centerline{\bf of cold dense matter}
\vskip 2.0cm
\centerline{P. Haensel
\footnote*{Permanent address: N. Copernicus Astronomical Center, Polish
Academy of Sciences, Bartycka 18, PL-00-716 Warszawa, Poland}
{}~~~~and ~~~~B. Pichon}
\vskip 0.5cm
\centerline{D.A.R.C. - U.P.R. 176 du C.N.R.S., Observatoire de Paris,
Section de Meudon}
\centerline{F-92195 Meudon Cedex, France}
\vskip 4cm
\parindent 0pt
Running title: Ground state of cold dense matter\par
Send proofs to: B. Pichon\par
Send offprint requests to: B. Pichon\par
Thesaurus: 04.02.1, 19.27.1, 19.50.1 \par
Section: 6. Formation, structure and evolution of stars\par
Submitted to: Main Journal\par
\vfill
\eject
\null\vskip 2cm
\baselineskip=22pt
\lineskip=2\lineskip
\lineskiplimit=2\lineskiplimit
\parindent 0pt
\vskip 1.0cm
{\bf Abstract.} We study the consequences of recent progress in the
experimental determination of masses of neutron rich nuclei for our
knowledge of the ground state of cold dense matter.  The most recent
experimental data determine the ground state of cold dense matter up to
$\rho \simeq 10^{11}~{\rm g~cm^{-3}}$.  The composition and the equation
of state of the ground state of matter, in this density interval, are
calculated.
\vfill
\eject
\parindent=20pt
\parindent=0pt
\centerline{\bf 1. Introduction }
\vskip 0.5cm
\parindent=20pt
Neutron stars are believed to be born as very hot objects, with initial
temperature exceeding $10^{11}$K.  At such temperatures matter can be
considered as being in a complete thermodynamic equilibrium (catalyzed
matter).  This approximation is usually extended to further stages of
evolution of neutron stars, where the temperature effect on stellar
structure can be neglected.  Standard neutron star models assume that
stellar matter is in its ground state (cold catalyzed matter) (see,
e.g., Shapiro \& Teukolsky 1983).
\par
The ground state of matter at the densities and pressures, at which all
neutrons are bound in nuclei (i.e.  below the neutron drip point) can be
described by a model formulated in the classical paper of Baym et al.
(1971, hereafter referred to as BPS).  An essential input for this model
are the ground-state masses of atomic nuclei, present in the lattice
sites of the crystal, and immersed in an electron gas.  At lowest
densities, the relevant nuclei are those, whose ground-state masses are
determined with high precision by the laboratory measurements.  However,
at higher densities the nuclei in the ground state of matter become more
and more neutron rich.  At the time, when the BPS paper was written, the
last experimentally studied nucleus, present in the ground state of
dense matter, was \nuc.Se.84. ($Z/A=0.405$), and the maximum density, at
which experimentally studied nucleus was present, was found to be
$\rho^{\rm exp}_{\rm max}= 8.2~10^{9}~{\rm g~cm^{-3}}$.
\par
During last two decades significant progress has been made in the
experimental nuclear physics techniques, and masses of many new neutron
rich isotopes have been measured (Wapstra \& Audi (1985);  latest
up-to-date results, used in the present work, have been evaluated by
Audi (1992, 1993)).  As we will show, the present experimental knowledge
of the masses of neutron rich atomic nuclei enables determination of the
composition and equation of state of the ground state of dense matter up
to about $\rho^{\rm exp}_{\rm max} \simeq 10^{11}~{\rm g~cm^{-3}}$.
\par
The model of dense matter is briefly described in Sect.  2.  Ground
state nuclear masses - available experimentally or evaluated via recent
mass formulae are discussed in Sect.  3.  Numerical results for the
composition and the equation of state of cold catalyzed matter are
presented in Sect.  4.  We also discuss there possible effects of
experimental uncertainties.  Sect.  5 contains discussion of our
results, with particular emphasis on the comparison between the results
based on the experimental data and those based on theoretical
calculations.  Finally, Sect.  6 presents our conclusions.
\vskip 1cm
\parindent=0pt
{\bf 2. Model of dense matter}
\parindent=20pt
\vskip 0.5cm
We shall assume that matter is in its ground state (complete
thermodynamic equilibrium - cold catalyzed matter) and that it forms a
perfect crystal with a single nuclear species, $(A,~Z)$, at lattice
sites.  Our formalism is similar as in BPS, with some modifications.
At given pressure $P$ the
equilibrium value of $A, Z$ is determined from the condition that the
Gibbs energy per nucleon be minimum.  The calculation is done using the
Wigner-Seitz (W-S) approximation.  The Gibbs energy of the W-S cell is
given by
$$ G_{cell}(A,Z)=W_N(A,Z)+W_L(Z,n_N)+[\epsilon_e(n_e,Z)+P]/n_N
\eqno(1) $$
where $W_N$ is the energy of the nucleus (including rest energy of
nucleons), $W_L$ is the body-centered cubic lattice energy per cell
(BPS), and $\epsilon_e$ is the mean
electron energy density.  The Gibbs energy per nucleon
$g=G_{cell}/A$.
\par
There are some differences between our expression for
$G_{cell}$, and the  BPS one. Our values of $W_N$ have been
obtained from the atomic masses by substracting not only the
 electron rest energies, but also removing the atomic electron
binding energies. Atomic binding energies were kept in the BPS
definition of $W_N$, to simulate the electron screening effects
in dense matter. In contrast to BPS, we use a much better
approximation for the electron screening effects  in dense
matter. Our expression for $\epsilon_e$ takes
into account deviations of the electron density from uniformity,
which result from the electron screening effects. We include
also the exchange term in $\epsilon_e$, which was neglected in
BPS.  For $\rho \gg 10^{6}~{\rm g~cm^{-3}}$ electrons are
ultrarelativistic, and our formula for $\epsilon_e$ becomes
$\epsilon_e(n_e,Z)=(1.00116
-1.78\times10^{-5}Z^{4/3})\epsilon_e^{\rm FG}(n_e)$, where
$\epsilon_e^{\rm FG}(n_e)$ is the energy density of an uniform,
free Fermi gas (Salpeter 1961).
\par
At given pressure, the values of electron density, $n_e$, and number
density of nuclei, $n_N$, are determined from the relations
$$ \eqalign{ n_e&=Zn_N~,\cr P&=P_e(n_e,Z)+P_L(n_N,Z)~.\cr }\eqno(2)$$
At the pressure $P_i$ at which optimal values $A, Z$ change into $A',
Z'$, matter undergoes a baryon density jump, $\Delta n_{\rm b}$, which
to a very good approximation is given by the formula
$$ {{\Delta n_{\rm b}\over n_{\rm b}}\cong {{Z\over A} {A'\over Z'} -1}}
\eqno(3) $$
The above equation results from the continuity of pressure, which is to
a very good approximation equal to the electron pressure, $P_e$.
\par
Actually, sharp discontinuity in $n_{\rm b}$ is a consequence of the
assumed one component plasma model.  Detailed calculations of the ground
state of dense matter by Jog $\&$ Smith (1982) show, that the transition
between the $A,~Z$ and $A',~Z'$ shells takes places through a very thin
layer of a mixed lattice of these two species.  However, since the
pressure interval within which the mixed phase exists is typically
$\sim 10^{-4}P_i$, the approximation of a sharp density jump is quite a good
representation of a nuclear composition of the ground state of matter.
\vskip 1cm
\parindent 0pt
{\bf 3. Nuclear masses}
\parindent 20pt
\vskip 0.5cm
Experimental masses of nuclei were taken from a recent evaluation of
Audi (1992, 1993).  Because of the pairing effect, only even-even nuclei
are relevant for the ground state problem.  In Fig.  1 we show the
network of all available experimental masses of even-even nuclei,
represented by filled circles. For the remaining isotopes,
up to the last one stable with respect to emission of a neutron (proton)
pair, we have used theoretical masses obtained with recent mass formula
(M\"oller (1992);  the description of the formalism can be found in
M\"oller \& Nix (1988)).  These nuclides are represented by open
circles.
\vskip 1cm
\parindent 0pt
{\bf 4. Numerical results}
\vskip 0.3cm
{\sl 4.1. Results obtained using mean experimental masses}
\parindent 20pt
\vskip 0.5cm
The baseline calculation has been done using the (mean) experimental
atomic masses, i.e.  without considering experimental errors.  As it
will be discussed in Sect.  4.2, these errors may become substantial for
nuclei with largest neutron excess.  The equilibrium nuclides present in
the cold catalyzed matter are listed in Table 1.  In the fifth column we
give the maximum density at which a given nuclide is present, $\rho_{\rm
max}$.  In the sixth column we give the value of the electron chemical
potential, $\mu_e$, at the density $\rho_{\rm max}$.  The transition to
the next nuclide has a character of a first order phase transition and
is accompanied by a density jump.  The corresponding fractional
increase of density, $\Delta\rho/\rho$,
is shown in the last column of Table 1.
The last row above the horizontal line, dividing the table into two
parts, corresponds to the maximum density, at which the ground state of
dense matter is determined by present up-to-date experimental data,
$\rho_{\rm max}^{\rm exp}$.  The table could be extended to higher
densities only by using theoretical determination of nuclear masses.  We
used the mass predictions of M\"oller (1992).  The very last line of
Table 1 corresponds to the neutron drip point in the ground state of
dense cold matter.  This limiting density can be determined exclusively
by the theoretical calculation.
\par
The values of $N$ and $Z$, corresponding to the ground state of dense
matter, are indicated in Fig.1 by large crosses.
\par
The last "experimental" row of Table 1 deserves a comment.  The nuclide
there is the most neutron rich isotope of nickel, whose mass has been
evaluated from experimental data.  In view of this, the determination of
the ground state of matter had to involve the mass formula for the
calculation of the mass of the next nickel isotope, i.e., $^{80}{\rm
Ni}$.  However, as long as our result for the ground state does not
depend on the {\it theoretical} mass formula used in the vicinity of
$^{78}{\rm Ni}$, our result can be treated as reliable.  We checked this
by switching to another recent theoretical mass formula of Pearson and
collaborators (1992;  private communication, for the description of the
model see Aboussir et al.  (1992));  the experimentally known nuclides
in the ground state of dense matter did not change.
Therefore, the determination of the ground state nuclide in the last row
of Table 1 should be treated as based on experiment (as long as we
neglect the experimental errors, see below).  However, the determination
of the value of $\rho_{\rm max}^{\rm exp}$ depends on the mass formula
used, because the next nuclide in the ground state of dense matter turns
out to be \nuc.Ru.126.  , sufficiently far from the most neutron rich
isotope of Ru experimentally available, \nuc.Ru.116.  , that its mass
depends in a nonnegligible way on the mass formula used for the
extrapolation.  In view of this, the pressure (and density) at which the
transition from \nuc.Ni.78.  to \nuc.Ru.126.  takes place depends, to
some extent, on the mass formula used to the extrapolation from the
experimental nuclear masses (see Tables 1, 2).  We get $\rho_{\rm
max}^{\rm exp}= 9.64~10^{10}~{\rm g~cm^{-3}}$ for the M\"oller (1992)
mass formula (Table 1), and $\rho_{\rm max}^{\rm exp}=1.08~10^{11}~{\rm
g~cm^{-3}}$ for the mass formula of Pearson and collaborators (1992)
(Table 2).
\par
The effect of the closed proton and neutron shells on the composition of
the ground state of matter is very strong ;  except for the \nuc.Fe.56.
nucleus, present in the ground state at lowest densities, all nuclides,
whose ground state mass is experimentally known, are those with a closed
proton or neutron shell (Table 1, Fig.  1).  A sequence of three
increasingly neutron rich isotopes of nickel $Z=28$ is followed by a
sequence of $N=50$ isotones of decreasing $Z$, ending at the last
experimentally measured \nuc.Ni.78..  This last nuclide is doubly magic
($N=50, ~Z=28$).
As we will discuss in the next section, this strong evidence for the
effect of closure of the $N=50$ shell is, at highest densities
considered, somewhat weakened when the experimental uncertainties are
taken into account.
\par
The equation of state constitutes an essential input for the calculation
of the neutron star structure.  In Table 3 we give equation of state for
the ground state of matter for $\rho < \rho_{\rm max}^{\rm exp}$.  It
has been calculated using the mean experimental masses of atomic nuclei.
For the sake of an easier comparison, we used the same pressure grid as
that used by BPS (except for the last line, which corresponds to
$\rho_{\rm max}^{\rm exp}$).  Generally, our "experimental" equation of
state is very similar to the BPS one.  However, in several density
intervals one notices a few percent difference, resulting from the
difference in the nuclides present at these densities.
\par
At the density exceeding $\rho^{\rm exp}_{\rm max}$, and up to the
neutron drip density, we get a sequence of $N=82$ isotones, of
decreasing proton number, from $Z=44$ down to $Z=36$ for the mass
formula of M\"oller (1992), with neutron drip at $\rho_{\rm
ND}=4.3~10^{11}~{\rm g~cm^{-3}}$ (Table 1).  The results obtained using
the theoretical mass formula of Pearson and collaborators (1992) (Table
2) are quite similar to those obtained using the mass formula of
M\"oller (1992), the differences reducing to two details:  the last
nucleus before neutron drip point has $Z=38$ (instead of $Z=36$ for
M\"oller (1992)), and the neutron drip takes place at a somewhat lower
density, $\rho_{\rm ND}=4.1~10^{11}~{\rm g~cm^{-3}}$.
\par
More detailed discussion of the importance of the shell terms in various
mass formulae, and in particular their comparison with experimental
evidence, will be presented in \S 5.
\par
The ground state composition at given pressure corresponds to the
absolute minimum of the Gibbs energy per nucleon, $g$, in the $N-Z$
plane.  Typically, there is only one well distinguished minimum (Fig.
2);  only close to the transition pressure between the two nuclear
species a well pronounced second minimum appears, and with inreasing
pressure becomes a new absolute minimum.  As seen in Fig.  2, the
absolute minimum lies in a valley, which may be called a "beta stability
valley" in superdense matter.  With increasing pressure, the valley
shifts in the $N$ direction, with only a slight change of the
inclination angle (Fig.  3).
\vskip 1cm
\parindent 0pt
{\sl 4.2.  Implications of experimental uncertainties}
\parindent 20pt
\vskip 0.5cm
Typical errors in experimentally measured masses of nuclei near the
normal (vacuum) beta stability valley are so small (at most a few tens
of \keV), that they are insignificant for the determination of the value
of $g$ :  they imply "experimental" uncertainty in $g$ of the order of
tens of \eV.  However, neutron excess in the nuclei present in the
ground state of dense matter increases with density.  At $\rho=2\times
10^{9}~{\rm g~cm^{-3}}$ the minimum of $g$ corresponds to \nuc.Kr.86.  ;
the experimental mass of this nucleus is known within $(\Delta W)_{\rm
exp}= 5 ~\keV $, which corresponds to an insignificant uncertainty
$(\Delta g)_{\rm exp}= 60~\eV$.  At $\rho \simeq 2\times 10^{10}~{\rm
g~cm^{-3}}$, the ground state nucleus turns out to be \nuc.Ge.82., with
$(\Delta W)_{\rm exp}= 155 ~\keV $, which implies $(\Delta g)_{\rm exp}=
2 ~\keV $.  Still, this uncertainty is not important for the
determination of the ground state of matter.  However, as one moves to
still higher pressure, the experimental uncertainties in $W$ become more
important.  We found, that of particular importance are large
uncertainties in the experimental masses of \nuc.Ni.76.  , \nuc.Ni.78.
and \nuc.Zn.82..
In what follows, we quote the results of Audi (1993).  The latest {\it
mean} experimental atomic mass excesses for $^{76}{\rm Ni}$, $^{78}{\rm
Ni}$, $^{80}{\rm Zn}$ and $^{82}{\rm Zn}$ are, respectively, $-42500$
keV, $-34920$ keV, $-51780$ keV, and $-42410$ keV.  The experimental
errors are:  $\Delta W(\nuc.Zn.82.)= 400 ~\keV $, $\Delta
W(\nuc.Ni.76.)= 900~ \keV $, $\Delta W(\nuc.Zn.80.)=170~ \keV$, and
$\Delta W(\nuc.Ni.78.)= 1100 ~\keV $.  These experimental uncertainties
determine upper and lower experimental bounds on $W$:  $W_{\rm
u.b.}=W+\Delta W$, and $W_{\rm l.b.}=W-\Delta W$.  In order to visualize
the possible effect of the experimental errors, we performed the
calculations of the ground state composition by replacing the masses of
$^{76}{\rm Ni}$ and $^{82}{\rm Zn}$ by their lower bounds, and the
masses of $^{78}{\rm Ni}$ and $^{80}{\rm Zn}$ by the corresponding upper
bounds.  The resulting modified lines of Table 1 are shown in the upper
part of Table 4.  Then we repeated the calculations, replacing masses of
$^{76}{\rm Ni}$, $^{78}{\rm Ni}$, $^{80}{\rm Zn}$ by their upper bounds,
and that of $^{82}{\rm Zn}$ by its lower bound.  Modified lines of Table
1 are displayed in this case in the lower part of Table 4.
 Both sets of modifications of  atomic masses are unfavorable for the
appearance of the doubly magic $^{78}{\rm Ni}$ in the ground
state of dense matter.
\par
Depending of our choice of experimentally allowed masses for the set
$^{76}{\rm Ni}$, $^{78}{\rm Ni}$, $^{80}{\rm Zn}$, $^{82}{\rm Zn}$, a
specific sequence of increasingly neutron rich nuclides closes the
experimental region of the ground state of dense matter.
\par
The results of Table 4 reflect the fact, that the valley in $g(N,~Z)$ in
the vicinity of $g_{\rm min}$ (Fig.  2, 3) is flat.  A specific example
of the immediate neighbourhood of the ground state minimum is shown in
Fig.  4.  Actually, by allowing the uncertainty $\pm (\Delta W)_{\rm
exp}$ we introduce the uncertainty (at a given pressure) in the ground
state values of $Z$ and $N$.  Within experimental uncertainties, the
ground state experimental nuclei at $\rho \gta 5\times 10^{10}~{\rm
g~cm^{-3}}$ are then no longer restricted to those with the $N=50$
closed shell.  For some experimentally allowed combinations of nuclear
masses, $N=48$ or $N=52$ appear in a density interval (Table 4).
For both combinations of atomic masses, the first
non-experimental nuclide appears at a somewhat lower density,
than in the case when mean experimental atomic masses were used.
Also, the first non-experimental $N=82$ nuclide is that with
$Z=46$, instead of $Z=44$. This is a consequence of using an
upper bound for the atomic mass of $^{78}{\rm Ni}$.
\par
The uncertainty in the "experimental" ground state composition of matter
at highest densities implies a corresponding uncertainty in the equation
of state.  In the density intervals, in which the ground state nuclide
is uncertain, the uncertainty in the "experimental" equation of state
may reach a few percent.
\vskip 1cm
\parindent 0pt
{\bf 5. Discussion }
\parindent 20pt
\vskip 0.5cm
In the present paper we calculated the composition and the equation of
state of the ground state of dense matter, using the most recent
experimental data on masses of very neutron rich nuclei.  The
experimental data determine the composition of the ground state of dense
matter up to $\rho_{\rm max}^{\rm exp}\simeq 10^{11}~{\rm g~cm^{-3}}$.
This conclusion does not depend on the particular choice of theoretical
determination of nuclear masses via the most recent mass formulae,
outside experimentally available region.
\par
If we vary nuclear masses within their experimental bounds, the
persistence of the magic number $N=50$ in the experimentally determined
ground state at $\rho \gta 5\times 10^{10}~{\rm g~cm^{-3}}$ becomes
somewhat weakened.  Had we used nuclear masses determined via
theoretical mass formulae (fitted to some earlier experimental data),
the strong effect of the closed $N=50$ shell would leave no possibility
for the appearence of the $N=48$, 52 nuclei in this density interval.
\par
By comparing experimental data for the \nuc.Ni.76., \nuc.Ni.76.,
\nuc.Zn.80.  or \nuc.Zn.82.  masses with their evaluation via
theoretical mass formulae, one notices a systematic overbinding of the
$N=50$ isotopes with respect to their neighbours, and with respect to
experiment (see Table 5).  The noticeable exception from this trend are
results of Pearson and collaborators (1992) and those given by the model
of M\"oller \& Nix (1988).
\par
Summarizing, both the calculation of the composition of the ground state
of dense cold matter, and the direct comparison with experimental atomic
masses indicate, that the additional binding resulting from the closure
of the $N=50$ shell for $Z/A\simeq 0.36 - 0.39$ nuclei is somewhat
weaker than that resulting from the shell term in most of the
theoretical mass formulae.
\par
While the persistence of the $N=50$ and/or $Z=28$ nuclei at lower
density may be treated as an {\it experimental fact}, the strong effect
of the $N=82$, dominating at $\rho_{\rm max}^{\rm exp}<\rho<\rho_{\rm
ND}$, might be an artifact of the extrapolation via the semiempirical
mass formulae.  Some many body calculations of the masses of very
neutron rich nuclei suggest, that the effect of the closed $N=82$ shell
might be much weaker, and could be replaced by the strong effect of the
closure of the $Z=40$ subshell (Haensel, Zdunik \& Dobaczewski 1989).
\vskip 1cm
\parindent 0pt
{\bf 6. Conclusion}
\parindent 20pt
\vskip 0.5cm
Using latest evaluation of experimental atomic masses of very neutron
rich nuclei, we were able to determine the ground state of cold dense
matter for the density up to $\simeq 10^{11}~{\rm g~cm^{-3}}$.
 By varying, within the present day experimental uncertainties, the
masses of nuclei in the vicinity of $Z=28$, $N=50$, we were able to
show, that the effect of the closure of the $N=50$ shell could be, at
$Z/A\simeq 0.36 - 0.39$, somewhat weaker than that obtained from most of
existing theoretical mass formulae.  We hope, that the future progress
in the mass evaluation of the neutron rich nuclei, will eventually yield
more precise experimental information about the actual effect of the
very large neutron excess on the additional binding due to the closure
of neutron shells.  The knowledge of this effect turns out to be
important for the determination of the ground state of cold, dense
matter.  This may seem to be of mostly academic interest, because of
expected deviations of the real neutron star crust from the ground
state.  However, this effect might be important also for the detailed
astrophysical scenarios of the cosmic nucleosynthesis of heavy nuclei.
\vskip 0.5cm
\parindent 0pt
{\it Acknowledgement.}
P.  Haensel has been supported in part by the Polish Committee for
Scientific Research (KBN), grant No.  2-1244-91-01.  The authors are
very grateful to Georges Audi to provide them with the up-to-date
results of his mass evaluation, and for his valuable advice.
\vfill
\eject
\centerline{\bf References}
\vskip 0.5cm
\parindent 0pt
\parskip=7pt
Aboussir, Y., Pearson, J.M., Dutta, A.K. \& Tondeur, F., 1992,
Nucl. Phys. A 549, 155 and references therein for previous papers
\par
Audi, G., 1992, Private communication
\par
Audi, G., 1993, Private communication
\par
Baym, G., Pethick, C., Sutherland, P., 1971, ApJ 170, 299
\par
Haensel, P. \& Zdunik, J.L., 1989, A \& A 229, 117
\par
Haensel, P., Zdunik, J.L., \& Dobaczewski, J., 1989, A \& A 222, 353
\par
Jog, C.J., Smith, R.A., 1982, ApJ 253, 839
\par
M\"oller, P., 1992, Unpublished, data available on the request from the
author
\par
M\"oller, P., \& Nix, J.R., 1988,  Atom. Data Nucl. Data Tables 39, 213
\par
Pearson, M., 1992, Private communication
\par
Salpeter, E.E., 1961, ApJ 134, 669
\par
\hangindent=1cm\hangafter=1
Shapiro, S.L., Teukolsky, S.A., 1983, Black Holes, White Dwarfs
and Neutron Stars, Wiley and Sons, New York
\par
Wapstra, A.H. and Audi, G., 1985, Nucl. Phys. A 432, 1
\par
\vfill
\eject
\centerline{\bf Figure caption}
Fig.1.  The network of even-even nuclei used in the present calculation.
Filled circles - experimental masses;  open circles - values
extrapolated using mass formula of M\"oller (1992).  The nuclides
present in the ground state of matter are indicated by crosses.
\vskip 0.5cm
Fig.2.  Map of $g-g_{min}$ in the $N - Z$ plane, for even - even nuclei,
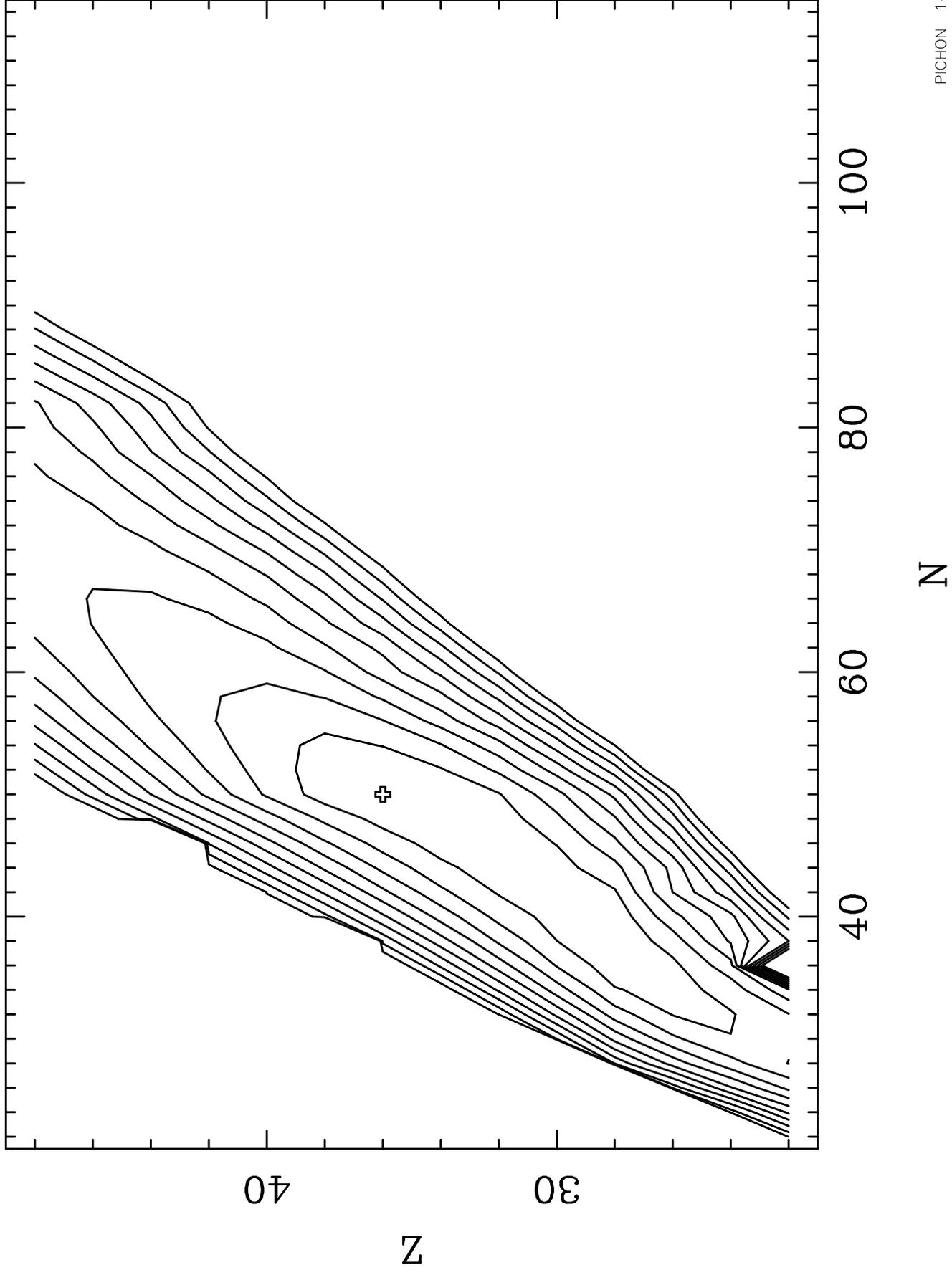
at $P=1.27~10^{27}~{\rm dyn~cm^{-2}}$.  The distance between contour
lines is 50 keV.  The absolute minimum corresponding to the ground state
is indicated by a cross.
\par
Fig.3.  Map of $g-g_{min}$ in the $N - Z$ plane, for even - even nuclei,
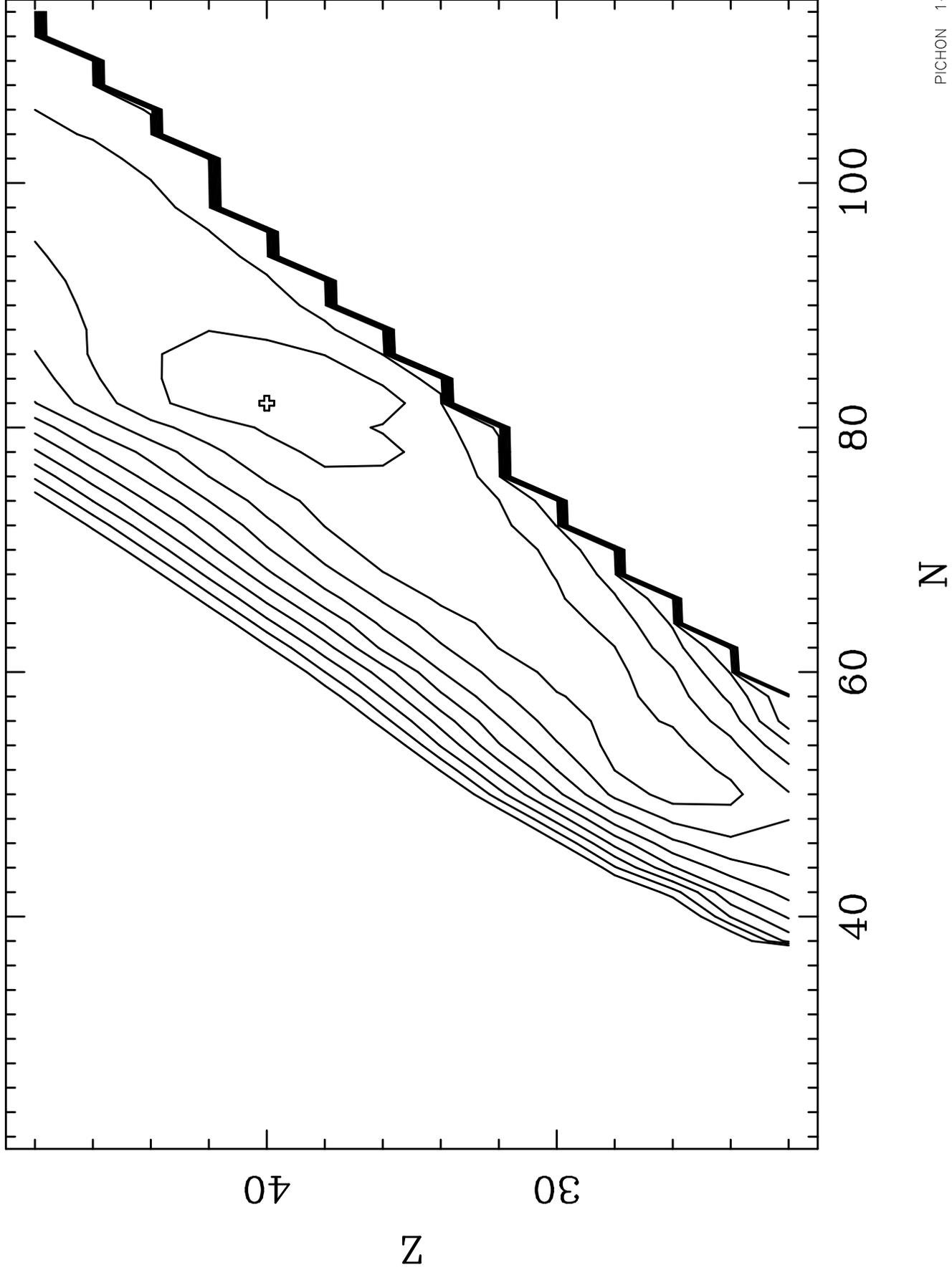
at $P=4.4~10^{29}~{\rm dyn~cm^{-2}}$.  Notation as in Fig.  2.
\par
Fig.4.  The excess of $G_{cell}$ (in keV) with respect to that
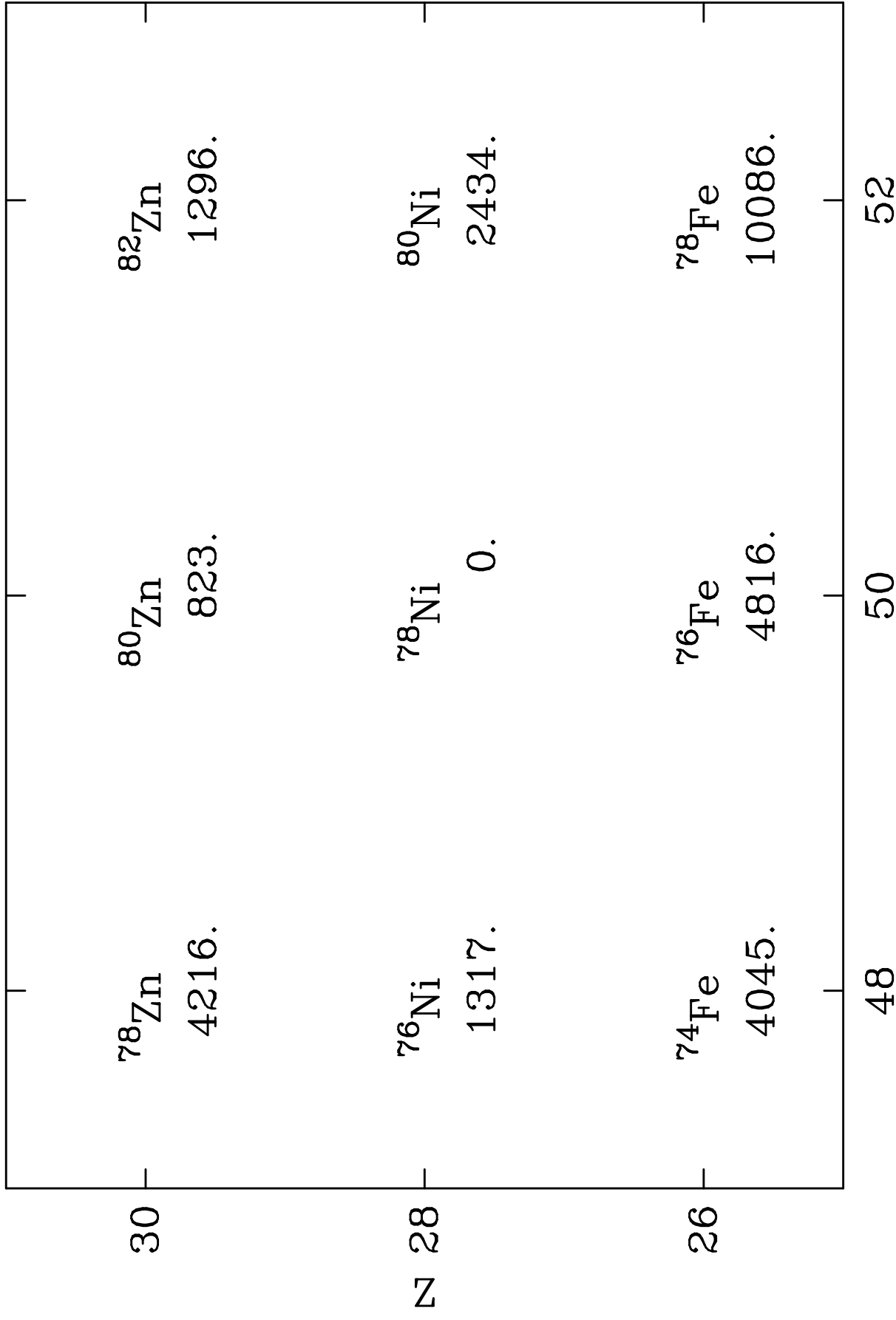
corresponding to the absolute minimum at $P=8~10^{28}~{\rm
dyn~cm^{-2}}$, in the immediate neighbourhood of the minimum.
\vfill
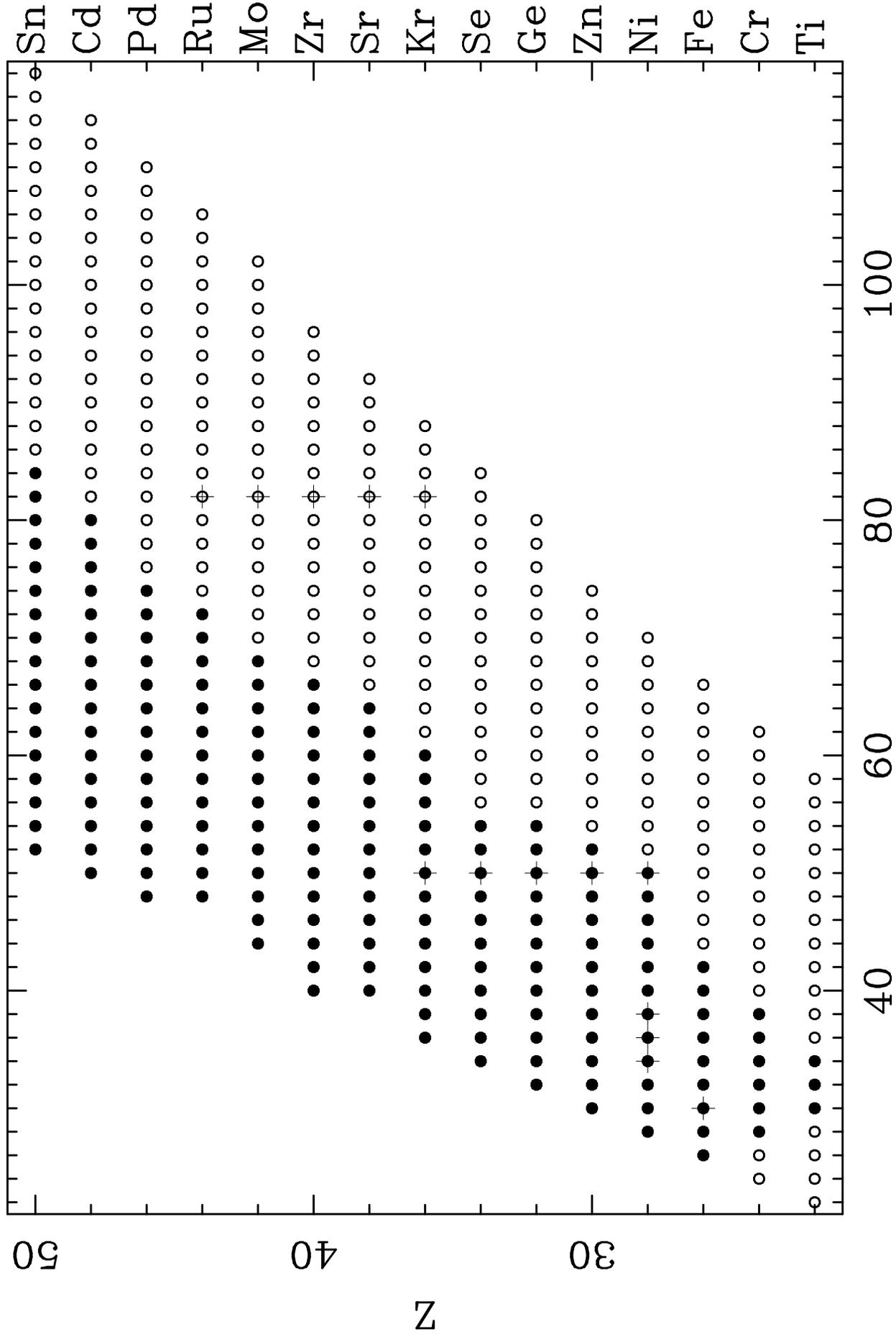
\end